\documentclass{iopjournal}

\usepackage[T1]{fontenc}
\usepackage[utf8]{inputenc}
\usepackage[numbers,square,sort&compress]{natbib}
\usepackage{braket}
\usepackage{amsmath}
\usepackage{graphicx}
\begin{document}

\title{Photoelectron interferometry with spectrally shaped polychromatic infrared pulses}

\author{E. A. Boati$^1$, G. Arvidsson$^1$, M. Ammitzb\"oll $^1$, P. K. Maroju$^1$, C. L\'ev\^eque$^2$, R. Weissenbilder$^1$, V. Shiriaeva$^1$, H. Laurell$^1$, M. Li$^3$, H. Wang$^3$, C. Dittel$^{1,4,5}$, M. Canhota$^1$, C. Guo$^1$, R. Ta\"{i}eb$^2$, J. Caillat$^2$, R. J. Squibb$^6$, R. Feifel$^6$, M. Gisselbrecht$^1$, C. L. Arnold$^1$, S. Luo$^3$, A. L’Huillier$^1$, D. Busto$^1$}

\affil{$^1$Department of Physics, Lund University, Box 118, 221 00 Lund, Sweden}

\affil{$^2$Sorbonne Université, CNRS, Laboratoire de Chimie Physique-Matière et Rayonnement, LCPMR, 75005 Paris, France}

\affil{$^3$Institute of Atomic and Molecular Physics, Jilin University, Changchun, 130012, China}

\affil{$^4$Physikalisches Institut, Albert-Ludwigs-Universit{\"a}t Freiburg, Hermann-Herder-Stra{\ss}e 3, 79104 Freiburg, Germany}

\affil{$^5$EUCOR Centre for Quantum Science and Quantum Computing, Albert-Ludwigs-Universit{\"a}t Freiburg, Hermann-Herder-Stra{\ss}e 3, 79104 Freiburg, Germany}

\affil{$^6$Department of Physics, University of Gothenburg, Origovägen 6B, 412 96 Gothenburg, Sweden}

\email{edoardo\_alberto.boati@fysik.lu.se, david.busto@fysik.lu.se}


\begin{abstract}
\begin{justify}
Laser-assisted photoelectron interferometry is a cornerstone of attosecond science, first used to characterize attosecond pulse trains and later to study photoionization dynamics. Extending this method to spectrally shaped polychromatic infrared probe fields enables encoding of information across multiple interferometric pathways within the photoelectron spectrum. Here, we experimentally demonstrate laser-assisted photoelectron interferometry using a spectrally shaped polychromatic infrared probe field composed of five distinct spectral components forming a Golomb ruler in the frequency domain. The measured interferograms exhibit multiple beating frequencies that agree with theoretical calculations, demonstrating the simultaneous encoding of multiple laser-assisted quantum beats in a single measurement. A quantitative analysis of the beating amplitudes shows that the strongly modulated temporal profile of the polychromatic probe introduces intensity- and delay-dependent distortions of the quantum beats that cannot be explained by second-order perturbation theory. These results establish the conditions required for the quantitative interpretation of polychromatic photoelectron interferometry and highlight the opportunities offered by spectro-temporal engineering of the probe field for future developments in attosecond science.
\end{justify}

\end{abstract}

\section{Introduction}
\begin{justify}
Photoionization is a fundamental process in light-matter interaction in which an electron, referred to as a photoelectron, is emitted from matter upon the absorption of a high-energy photon. It constitutes the underlying mechanism of a broad range of photoelectron-based measurement techniques, including spectroscopy~\cite{Becker1995}, microscopy~\cite{Dabrowski2020}, tomography~\cite{Bourassin2020,laurellNatPhot2025,LaurellSR2025,Morrigan2023}, and interferometry~\cite{Veniard1996,Wollenhaupt2002,You2020,Maroju2023}. Many of these techniques rely on ultrashort light pulses to induce photoionization, thereby generating an electron wavepacket (EWP). Since short light pulses consist of a coherent superposition of a broad range of frequencies, the emitted photoelectron is naturally described as a coherent superposition of continuum states at different energies with well-defined relative phases, forming a coherent wavepacket that evolves in time.

Photoelectron interferometry uses pulses with different central frequencies to induce interference among multiple ionization pathways that lead to the same final state, thereby characterizing and coherently controlling the EWPs. 
This technique has been implemented in various experimental schemes, including multiphoton ionization (MPI) \cite{Wollenhaupt2002,Beaulieu2017} and laser-assisted photoionization based on infrared (IR) and extreme-ultraviolet (XUV) light pulses \cite{Paul2001,Villeneuve2017,Cheng2020,Maroju2023}.
Photoelectron interferometry based on MPI [Fig.~\ref{fig:1}(a)] has been extensively investigated using shaped femtosecond laser pulses. In this context, pulse shaping provides a versatile means of controlling the coherent superposition of multiphoton ionization pathways. By tailoring the spectral amplitude, phase, and polarization of the ionizing fields, the relative amplitudes and phases of photoelectron partial waves can be manipulated, enabling coherent control over the emitted EWPs \cite{Kerbstadt2019}. This approach has been used to engineer complex photoelectron wavepackets with nontrivial angular structures \cite{Kohnke2023,Kohnke2026}, demonstrating the power of pulse shaping to manipulate interference between ionization pathways.

In contrast, XUV-IR photoelectron interferometry in attosecond science has primarily been developed as a metrological tool, and the opportunities opened by pulse shaping have not yet been fully explored. Reconstruction of attosecond beating by interference of two-photon transitions (RABBIT) is a fundamental technique in this field, used to characterize XUV pulse trains~\cite{Paul2001}. 
It relies on the generation of photoelectrons due to the absorption of a single photon from an XUV attosecond pulse train composed of odd-order harmonics, and the absorption or stimulated emission of an IR photon from the time-delayed fundamental IR field, leading to interference between different pathways as illustrated in Fig.~\ref{fig:1}(b).
By studying how the photoelectron spectrum evolves as a function of the time delay between the XUV and IR fields, information about the XUV field itself can be extracted~\cite{LopezMartens2005,Nayak2025}. Beyond attosecond pulse reconstruction, RABBIT has also become a powerful technique for investigating photoemission dynamics on its natural timescale~\cite{Klunder2011, Gruson2016,Zhong2020,Peschel2022,Cattaneo2018,Jordan2020,LiPRL2025,Makos2025,Locher2015}.

While RABBIT is useful for fully characterizing a coherent EWP, more advanced techniques are needed to characterize a photoelectron in a mixed quantum state. One of the first applications of IR pulse shaping in XUV-IR pump-probe photoelectron interferometry has been the development of a photoelectron quantum-state tomography protocol, KRAKEN, a Swedish acronym for Quantum State Tomography of Attosecond Electron Wavepackets \cite{Laurell2022,laurellNatPhot2025}. By combining an ultrashort XUV pulse with a tunable bichromatic IR probe field [Fig.~\ref{fig:1}(c)], KRAKEN extends conventional RABBIT interferometry and enables the reconstruction of the photoelectron density matrix.
The same approach was recently applied to free electron tomography in ultrafast electron microscopy \cite{FangArxiv2026}. Additionally, the use of trichromatic IR pulses in laser-assisted photoionization has been proposed as a platform for testing Born's rule of quantum mechanics \cite{Forderer2025}. These examples illustrate the growing interest in exploiting spectrally engineered polychromatic fields to extend the capabilities of laser-assisted photoelectron interferometry beyond those achievable with conventional monochromatic probe fields.

In our present work, we experimentally demonstrate laser-assisted photoelectron interferometry using a polychromatic IR field composed of five spectral components arranged as a Golomb ruler in the spectral domain \cite{Robinson1979}, thereby extending the capabilities of existing monochromatic or bichromatic schemes [see Fig.~\ref{fig:1}(d)]. We show that this approach enables the simultaneous measurement of multiple interferometric beatings. At the same time, the strongly modulated temporal intensity profile of the polychromatic IR probe pulse can introduce additional intensity- and delay-dependent effects that must be carefully considered for the quantitative interpretation of the measurements.

In Section~\ref{sec:2}, we introduce the principle of laser-assisted photoelectron interferometry using a polychromatic IR field. In Section~\ref{sec:3}, we describe the experimental setup, present the experimental results, and compare them with numerical calculations based on the strong-field approximation (SFA) and the time-dependent Schrödinger equation (TDSE). In Section~\ref{sec:4}, we analyze the results within the framework of second-order perturbation theory. In Section~\ref{sec:5}, we discuss how higher-order effects influence polychromatic photoelectron interferometry. 
\begin{figure}[htbp]
    \centering
    \includegraphics[width=\linewidth]{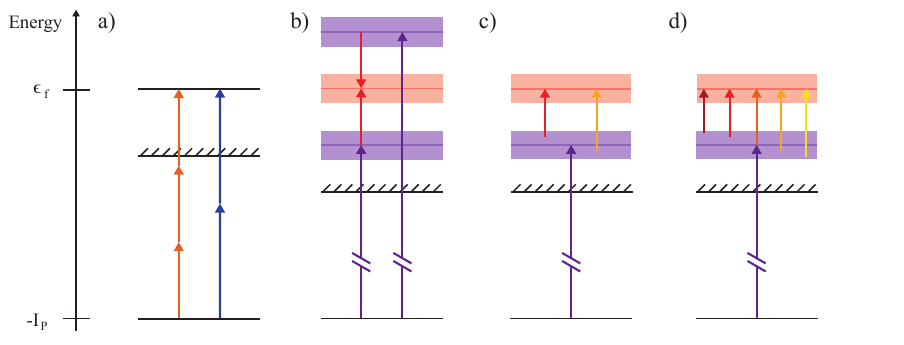}
    \caption{Energy diagrams showing different schemes of photoelectron interferometry. a) Bichromatic multiphoton ionization. b) The RABBIT scheme. c) The KRAKEN scheme. d) Polychromatic laser-assisted photoionization.}
    \label{fig:1}
\end{figure}
\end{justify}

\section{Laser-assisted photoelectron interferometry with polychromatic fields}
\label{sec:2}
\begin{justify}

Laser-assisted photoelectron interferometry relies on the generation of a photoelectron by absorption of an XUV photon in the presence of an IR laser field. Due to the short duration of the XUV pulse, a broad distribution of continuum states is populated. The IR laser field can induce transitions between continuum states via absorption or stimulated emission of IR photons, resulting in sidebands in the photoelectron spectrum. If both the XUV and IR pulses have broad spectra, multiple combinations of XUV and IR frequencies can couple different intermediate continuum states to the same final photoelectron energy $\epsilon_f$. Consequently, the measured signal at a given final energy contains contributions from many laser-assisted ionization pathways, making it difficult to isolate the contribution of each individual coherence of the photoelectron wavepacket created by absorption of the XUV light pulse only~\cite{JimenezGalan2016,Busto2018}. 

Alternatively, a spectrally shaped IR field composed of a few narrow spectral components can be used to control the number of ionization pathways leading to a given final state.
This concept is illustrated by the bichromatic KRAKEN scheme \cite{laurellNatPhot2025,Laurell2022}, where an IR field with two narrow spectral components $\omega_1$ and $\omega_2$ is used to induce a transition from two different intermediate continuum states $\ket{\epsilon_1}$ and $\ket{\epsilon_2}$ to the same final state $\ket{\epsilon_f}$ [Fig.~\ref{fig:1}(c)], with $\epsilon_1+\hbar\omega_1=\epsilon_2+\hbar\omega_2=\epsilon_f$. The interference of the corresponding ionization paths gives rise to oscillations in the photoelectron yield as a function of the time delay $\tau$ between the XUV and the bichromatic IR field. Within second-order perturbation theory, where the interaction consists of the absorption of a single XUV photon followed by the absorption or stimulated emission of a single IR photon, the photoelectron signal at the final energy $\epsilon_f$ is described by \cite{Laurell2022}
\begin{equation}
\begin{aligned}
S(\epsilon_f,\tau)
&\approx|E_1d_{\epsilon_f,\epsilon_1}|^2\bra{\epsilon_1}\rho\ket{\epsilon_1}
+|E_2d_{\epsilon_f,\epsilon_2}|^2\bra{\epsilon_2}\rho\ket{\epsilon_2} \\
&\quad
+E_1 E_2^* d_{\epsilon_f,\epsilon_1}  d^*_{\epsilon_f,\epsilon_2}
e^{i\delta\omega\tau}\bra{\epsilon_1}\rho\ket{\epsilon_2}
+E_2  E_1^* d_{\epsilon_f,\epsilon_2} d^*_{\epsilon_f,\epsilon_1}
e^{-i\delta\omega\tau}\bra{\epsilon_2}\rho\ket{\epsilon_1},
\end{aligned}
\label{eq:1}
\end{equation}
where $\braket{\epsilon_f|d|\epsilon_i}=d_{\epsilon_f,\epsilon_i}$ ($i=1,2$) are the continuum-continuum dipole transition matrix elements from $\ket{\epsilon_i}$ to $\ket{\epsilon_f}$ with $d$ the dipole operator, $E_i$ are the amplitudes of the IR spectral components, and $\rho$ is the density matrix describing the photoelectron wavepacket created by the absorption of a single XUV photon. The first two terms are independent of the time delay $\tau$ between the bichromatic IR pulse and the ionizing XUV pulse, and are proportional to the populations of the photoelectron density matrix. The last two terms give rise to oscillations as a function of the delay $\tau$, at a frequency $\delta\omega = \omega_1 - \omega_2$ equal to the difference between the two frequency components of the bichromatic IR field. For approximately constant dipole transition matrix elements, the amplitude of these oscillations is proportional to the off-diagonal element $\bra{\epsilon_1}\rho\ket{\epsilon_2}$, i.e., the coherence of the density matrix $\rho$. The frequency selectivity offered by the spectral shaping of the probe field therefore allows measuring the coherences of the photoelectron density matrix \cite{laurellNatPhot2025}.

The KRAKEN protocol employs a bichromatic IR probe field to measure a single coherence of the photoelectron density matrix at a time. It therefore relies on several measurements using different frequencies of the bichromatic field to obtain the complete density matrix. It can be more advantageous to use probe fields composed of multiple spectral components to obtain information on several coherences in a single measurement. The same perturbative treatment discussed above can be generalized to an IR field composed of $N$ spectral components, centered at frequencies $\omega_1$, ..., $\omega_N$, with amplitudes $E_1$, ..., $E_N$, coupling the intermediate continuum states $\ket{\epsilon_1}$, ..., $\ket{\epsilon_N}$ to the same final continuum state $\ket{\epsilon_f}$ as shown in Fig.~\ref{fig:1}(d).
In this case, the oscillations of the photoelectron yield as a function of the time delay $\tau$ are given by
\begin{equation}
S(\epsilon_f,\tau)\approx\sum_{n=1}^N |E_nd_{\epsilon_f,\epsilon_n}|^2\bra{\epsilon_n}\rho\ket{\epsilon_n}
+\sum_{\substack{m,n=1 \\ m\neq n}}^N E_m E_n^*d_{\epsilon_f,\epsilon_m}d_{\epsilon_f,\epsilon_n}^*e^{i\delta\omega_{mn}\tau}\bra{\epsilon_m}\rho\ket{\epsilon_n},
\label{Eq:PolyKRAKEN}
\end{equation}
where $\delta\omega_{mn}=\omega_m-\omega_n$. As above, the first term is proportional to the populations in the density matrix $\rho$, while the second term is proportional to the coherences. In principle, an IR field composed of $N$ spectral components gives rise to $N(N-1)/2$ pairwise interference terms. All of them can be independently evaluated from the Fourier spectrum, provided that each pair of IR frequencies gives rise to a unique beating frequency $\delta\omega_{mn}$. This condition is fulfilled by choosing the IR frequencies so that they form a so-called Golomb ruler in the spectral domain \cite{Robinson1979}. A Golomb ruler is a mathematical object defined as a set of ticks placed at integer positions along a ruler such that the distance between any two ticks is unique. In the spectral domain, this corresponds to a set of frequencies arranged so that, given the smallest spectral spacing $\delta\omega$, every pairwise frequency difference is a unique integer multiple of $\delta\omega$.

Equation~(\ref{Eq:PolyKRAKEN}) therefore provides a simple framework for interpreting laser-assisted photoelectron interferometry with spectrally shaped IR fields. It predicts that the delay-dependent photoelectron signal consists of a superposition of oscillations at all pairwise beating frequencies of the probe field, each directly related to a coherence of the photoelectron density matrix $\rho$. Consequently, the Fourier analysis of a single spectrogram, recorded by scanning the photoelectron signal as a function of the time delay $\tau$, offers the possibility to retrieve all of these coherence terms simultaneously.

\end{justify}

\section{Experimental and simulated results}
\label{sec:3}
\begin{justify}
\begin{figure*}
    \centering
    \includegraphics[width=\textwidth]{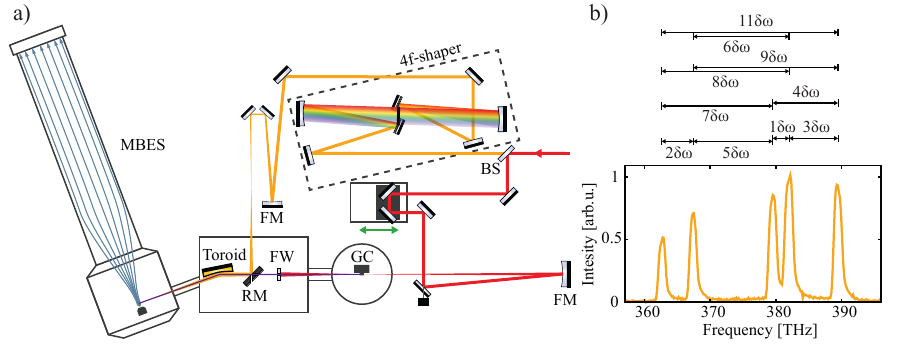}
    \caption{(a) Schematic representation of the experimental setup used for polychromatic photoelectron interferometry. The labeled components are BS, beam splitter; FM, focusing mirror; GC, gas cell; FW, filter wheel; and RM, recombination mirror. The green arrow denotes a motorized attosecond-scale delay stage. (b) Measured IR spectrum after the 4f-shaper. The pairwise frequency differences between the five transmitted spectral components are shown above the spectrum in units of $\delta \omega$.}
    \label{fig:2}
\end{figure*}

Figure~\ref{fig:2} presents a schematic representation of the experimental setup used in this work (details can be found in \cite{luo2023}). A Ti:Sapphire laser is used to generate 4\,mJ, 20\,fs IR pulses with a central wavelength of 800\,nm at a repetition rate of 3\,kHz. The pulses are sent into an actively stabilized Mach-Zehnder interferometer. The arm of the interferometer that is reflected off the beam splitter (50\% of the input energy) is equipped with a motorized delay stage and used to generate high-order harmonics by focusing the laser pulses into an argon-filled cell with a spherical mirror of 50\,cm focal length. After the generation cell, the co-propagating IR radiation is filtered out with a metallic filter that transmits XUV radiation while blocking the IR and low-order harmonics. Here, we use a combination of aluminum and germanium filters, which allows us to spectrally shape the generated XUV pulses, creating a sharp spectral edge in our XUV spectrum around 30\,eV. The highest harmonic order that is efficiently transmitted through the filter is the 19$^{\mathrm{th}}$.

The other arm of the interferometer is equipped with a 4f-spectral shaper. A first grating separates the spectral components, which are then focused at different spatial positions by a spherical mirror placed one focal length away from the grating. In the focal plane of the mirror, a mask composed of five slits is inserted. The slits are arranged so that the transmitted radiation corresponds to a Golomb ruler in the spectral domain. A second set of a focusing mirror and a grating is used to overlap the transmitted spectral components spatially and to ensure that the beam transmitted through the 4f-shaper is collimated and free from spatio-temporal couplings. Figure~\ref{fig:2} (b) shows the IR spectrum transmitted through the 4f-shaper with the mask inserted in the Fourier plane of the 4f-shaper. It consists of five frequencies forming a spectral Golomb ruler, as indicated on top of the spectrum. The resulting pulse is then recombined with the XUV pulse using a drilled recombination mirror, and both the XUV and IR pulses are focused into a helium gas jet using a gold-coated toroidal mirror. The interaction between the two pulses and the gas leads to the ionization of helium atoms. The kinetic energy of the emitted photoelectrons is measured using a 2-m-long magnetic bottle electron spectrometer with spectral resolution better than 100\,meV for electrons with kinetic energies below 10\,eV. 

\begin{figure}[htbp]
    \centering
 \includegraphics[width=\textwidth]{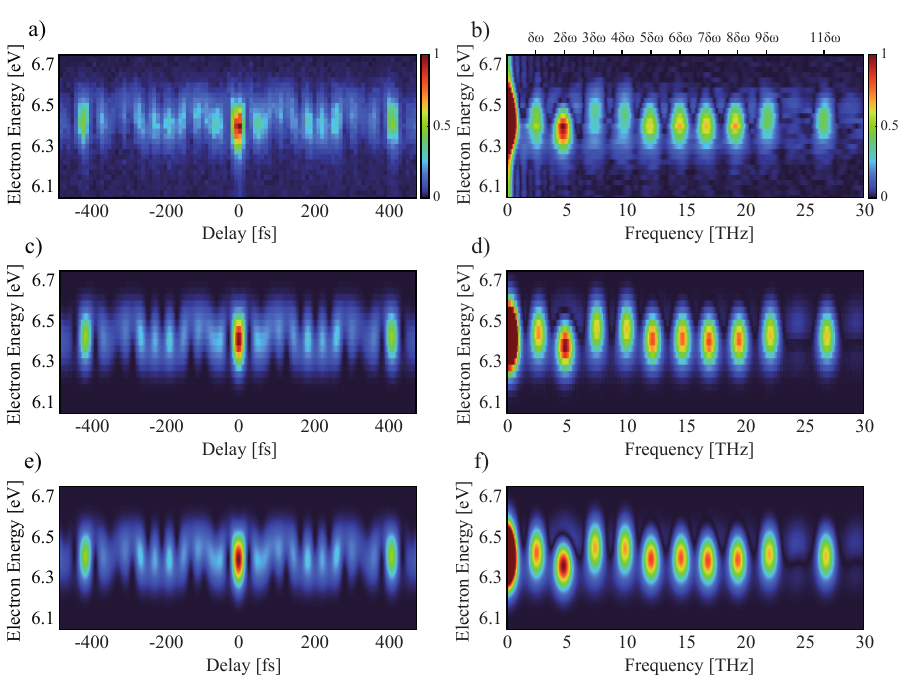}

    \caption{Comparison between experimental measurements (a,b), SFA simulations (c,d), and TDSE simulations (e,f) for polychromatic laser-assisted photoionization with five different IR spectral components. (a,c,e) Normalized photoelectron spectrograms. (b,d,f) Corresponding normalized Fourier maps obtained by Fourier transforming the spectrograms along the delay axis, normalized to the peak of the $2\delta\omega$ beating frequency. The beating frequencies are indicated on top of (b). A color bar is indicated on the right of (a) and (b).}
    \label{fig:3}
\end{figure}

Figure~\ref{fig:3}(a) presents the photoelectron spectrum as a function of the time delay between the XUV and IR pulses, for sideband 20, obtained by absorption of harmonic 19 together with an IR photon. The spectrogram shows a complex interference pattern with clear but irregular oscillations as a function of time delay. Additionally, the time-delay-dependent interference pattern varies with the photoelectron's final energy. Figure~\ref{fig:3}(b) shows the Fourier map obtained by computing the Fourier transform of the interferogram along the delay axis for each final energy. As predicted by Eq.~(\ref{Eq:PolyKRAKEN}), the Fourier spectrum consists of ten discrete beating frequencies corresponding to the ten unique pairwise frequency differences that result from the five spectral components forming the Golomb ruler. The absence of a component at 10\,$\delta\omega$ is a direct consequence of the chosen fifth-order Golomb ruler, i.e., a Golomb ruler with five ticks [see Fig.~\ref{fig:2}(b)]. The different beating frequencies are centered at different photoelectron energies since each of them originates from a distinct pair of IR spectral components, such that pairs involving higher frequencies (e.g., for $3\,\delta\omega$ and $4\,\delta\omega$) result in slightly higher electron energies as compared to pairs involving lower frequencies (e.g., $2\,\delta\omega$). These observations constitute the experimental signature of polychromatic laser-assisted photoelectron interferometry and demonstrate the simultaneous generation of multiple interferometric beatings within a single delay scan.

The experimental results are compared with numerical simulations based on the strong-field approximation (SFA) \cite{Lewenstein1994,Mairesse2005} and the time-dependent Schrödinger equation (TDSE). The SFA provides a simple framework for investigating the physical mechanisms underlying the measured spectrograms, while the TDSE provides a rigorous numerical solution of the light-matter interaction. Within the SFA, the amplitude of the photoelectron wavepacket is given in atomic units by
\begin{equation}
    b(\textbf{p},\tau)=\text{i}\int_{-\infty}^{+\infty} \ \textbf{E}_\mathrm{XUV}(t')\cdot\textbf{d}[\textbf{p}+\textbf{A}(t',\tau)]\exp\left\{\text{i}\int_{t'}^{+\infty } \frac{1}{2}\left[\textbf{p}+\textbf{A}(t'',\tau)\right]^2dt''+\text{i}I_pt'\right\}\mathrm{d}t', 
    \label{Eq:SFA}
\end{equation}
where $I_p$ is the ionization potential, $\textbf{p}$ is the canonical momentum,  $\textbf{A}$ the IR vector potential, $\textbf{E}_\text{XUV}$ is the XUV electric field and $\textbf{d}[\textbf{p}+\textbf{A}(t',\tau)]=\braket{\textbf{p}+\textbf{A}(t',\tau)|\textbf{d}|g}$ is the dipole matrix element for the transition from the ground state to the final state with momentum $\textbf{p}+\textbf{A}(t',\tau)$, which we assume to be constant over the energy range considered here. The simulations are performed along the (common) polarization direction of the XUV and IR fields. They are benchmarked against full TDSE calculations in atomic hydrogen in the velocity gauge. The wavefunction is expanded in spherical harmonics and represented radially on a finite-element discrete-variable representation \cite{Rescigno2000}. The photoelectron spectra are extracted with the time-dependent Surface Flux (t-SURFF) method \cite{Tao2012}, and infinite-range exterior complex scaling is used to avoid unphysical reflections \cite{Scrinzi2010}. The TDSE and SFA simulations use pulses with Gaussian profiles centered at the same frequencies as in the experiments. 
The different spectral components of the IR field are assumed to have the same intensity $I_0=2\times10^{10}$\,W/cm$^2$. The results of the calculations, shown in Fig.~\ref{fig:3}(c-f), are in excellent agreement with each other and reproduce the measured spectrograms and Fourier maps well, including the energy dependence of the beating frequencies and their relative amplitudes.
\end{justify}

\section{Analysis based on second-order perturbation theory}
\label{sec:4}
\begin{justify}

We analyze the results by fitting the experimental signal for each final energy using 
\begin{equation}
    S(\epsilon_f,\tau)=\exp\left[-\frac{\tau^2}{2\sigma_\tau^2 }\right]\left[A_0(\epsilon_f)+\sum_{n=1}^{11}B_n(\epsilon_f)\cos\left(n\delta\omega\tau-\varphi_n(\epsilon_f)\right)\right],
    \label{Eq:fit}
\end{equation}
where $\sigma_\tau$ parametrizes the temporal width of the XUV-IR cross-correlation trace, $A_0(\epsilon_f)$ is a constant background, $B_n(\epsilon_f)$ and $\varphi_n(\epsilon_f)$ are the amplitude and phase of the oscillations at frequency $n\delta\omega$, respectively, where we set $B_{10}(\epsilon_f)=0$. Figure~\ref{fig:4}(a) presents the fitted oscillation amplitudes (orange) as a function of the final energy. We observe that the amplitude varies between different beating frequencies, with the beating frequency $2\delta\omega$ showing the largest amplitude.  According to Eq.~(\ref{Eq:PolyKRAKEN}), the amplitude of the oscillations at frequency $\delta\omega_{mn}$ depends on five parameters: the amplitude of the two corresponding spectral components, $E_m$ and $E_n$, the amplitude of the corresponding dipole matrix elements $d_{\epsilon_f,\epsilon_m}$ and $d_{\epsilon_f,\epsilon_n}$, and the value of the density matrix element $\bra{\epsilon_m}\rho\ket{\epsilon_n}$. We now investigate how each of these quantities affects the measured oscillation amplitudes.

\begin{figure}[htbp]
    \centering
 \includegraphics[width=\textwidth]{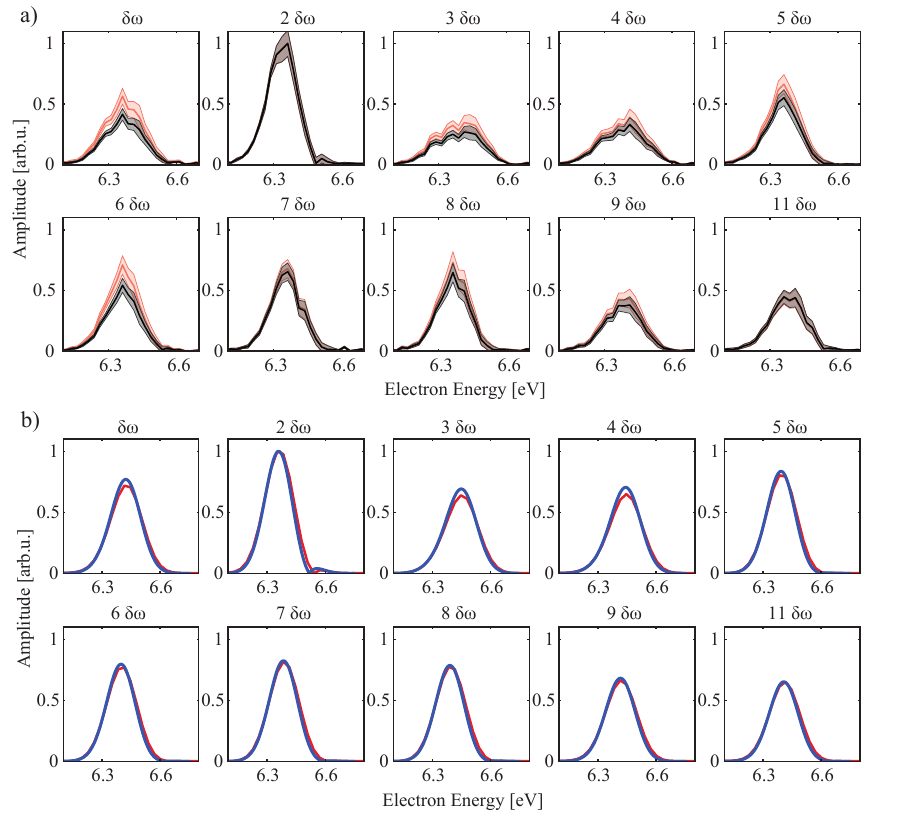}

    \caption{(a) Comparison between experimentally measured oscillation amplitudes for different beating frequencies before (orange) and after (grey) the normalization with respect to the strength of the IR spectral components. All the curves are normalized to the maximum of the $2\,\delta\omega$ oscillation amplitude. The shaded area indicates the standard deviation extracted from the fit.  (b) Oscillation amplitudes for different beating frequencies extracted from SFA (red) and TDSE (blue) calculations at an intensity of $2.00 \times 10^{10}$ W/cm$^2$ for each IR spectral component.}
    \label{fig:4}
\end{figure}

We first focus on the role of the relative amplitudes of the IR spectral components. Based on the measured IR spectrum [Fig.~\ref{fig:2}(b)], the beating frequencies $\delta\omega$, $3\,\delta\omega$, $4\,\delta\omega$, and $6\,\delta\omega$, which involve the strongest spectral components, would be expected to dominate the Fourier spectrum. However, the opposite trend is observed experimentally. Normalizing the fitted oscillation amplitudes by the product of the corresponding IR spectral amplitudes [grey data in Fig.~\ref{fig:4}(a)] has a very small impact on the amplitudes, demonstrating that differences in the strength of the IR spectral components cannot explain the observed modulation. This conclusion is further supported by the SFA and TDSE calculations [Fig.~\ref{fig:4}(b)], which show a similar behavior despite using IR spectral components with the same amplitudes.

Next, we focus on the role of the dipole matrix elements $d_{\epsilon_f,\epsilon_i}$. These matrix elements can be calculated analytically using the approximate expression \cite{Dahlstrom2012,ji2024}
\begin{equation}
d_{\epsilon_f,\epsilon_i}\approx\frac{e^{-\frac{\pi}{2}(\frac{1}{k_i}-\frac{1}{k_f})}(2k_i)^{\frac{i}{k_i}}\Gamma\left[2+i\left(\frac{1}{k_i}-\frac{1}{k_f}\right)\right]}{|k_i-k_f|^2(2k_f)^\frac{i}{k_f}(k_i-k_f)^{\left(\frac{i}{k_i}-\frac{i}{k_f}\right)}(k_ik_f)^{\frac{1}{2}}},
\label{eq:4}
\end{equation}
where we have used atomic units and where $k_i$ and $k_f$ are the momenta associated with photoelectrons of kinetic energy $\epsilon_i$ and $\epsilon_f$ respectively. $\Gamma(z)$ is the complex Gamma function. Figure~\ref{fig:5} shows the amplitude of the continuum-continuum transition matrix elements as a function of the IR wavelength, for two fixed final photoelectron kinetic energies. Changing the IR wavelength therefore corresponds to coupling a different intermediate continuum state to the same final state. For both kinetic energies, the transition strength increases smoothly with increasing wavelength, favoring transitions driven by the red-most spectral components of the Golomb ruler. This trend is consistent with the experimentally observed enhancement of the $2\,\delta\omega$ beating relative to the $\delta\omega$, $3\,\delta\omega$ and $4\,\delta\omega$ components. However, the wavelength dependence of the continuum-continuum matrix elements is too small to account for the measured variation. Across the probed spectral range, the product of the continuum-continuum matrix elements involved [see Eq.~(\ref{Eq:PolyKRAKEN})] varies at most by 20\%, whereas the oscillation amplitudes differ by approximately 40\% between, e.g., the $2\,\delta\omega$ and $3\,\delta\omega$ components. Although Eq.~(\ref{eq:4}) is approximate, the magnitude of the effect is in good agreement with many-body perturbation theory calculations \cite{laurellNatPhot2025}, indicating that continuum-continuum transition strengths alone cannot explain the observed modulations in the oscillation amplitudes.

\begin{figure}
    \centering
    \includegraphics[width=\textwidth]{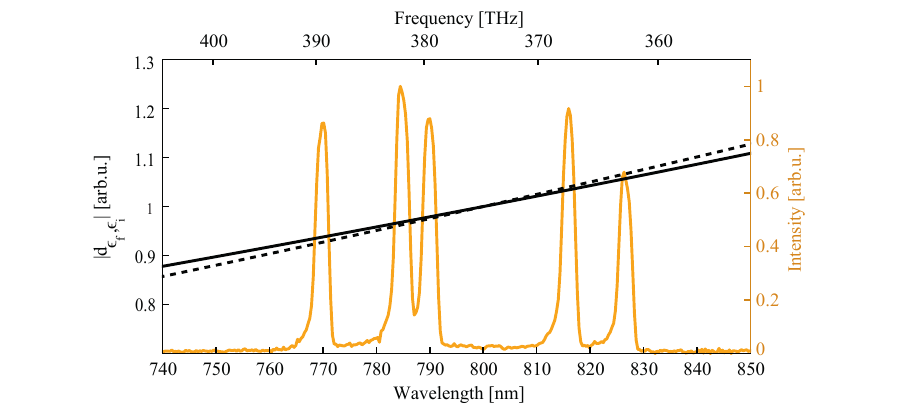}
    \caption{The solid and dashed curves show the dependence of the continuum-continuum dipole matrix elements on the IR wavelength for sidebands 20 and 32, respectively, corresponding to final photoelectron energies of 6.4 eV and 25.0 eV. The yellow curve displays the experimental IR spectrum [same as in Fig.~\ref{fig:2}(b)].}
    \label{fig:5}
\end{figure}

Finally, the last parameter in Eq.~(\ref{Eq:PolyKRAKEN}) that can affect the beating amplitudes is the density matrix element $\bra{\epsilon_m}\rho\ket{\epsilon_n}$. Describing the photoelectron as a pure state, which is justified for photoionization of helium in this energy range \cite{laurellNatPhot2025}, the coherences of the photoelectron density matrix are given by $|\bra{\epsilon_m}\rho\ket{\epsilon_n}|=\sqrt{\bra{\epsilon_m}\rho\ket{\epsilon_m}\bra{\epsilon_n}\rho\ket{\epsilon_n}}$, where the populations $\bra{\epsilon_m}\rho\ket{\epsilon_m}$ follow the Gaussian-like spectrum of harmonic 19. As a result, the coherences are expected to decrease smoothly as $\epsilon_m-\epsilon_n=\hbar\delta\omega_{mn}$ increases, leading to a corresponding smooth reduction of the oscillation amplitudes for increasing beating frequencies. Experimental decoherence, such as the finite energy resolution of the photoelectron spectrometer, would further limit the observation of coherences \cite{Bourassin2020,LaurellSR2025}, but would preserve the monotonic behavior. Consequently, the photoelectron density matrix cannot account for the pronounced modulation of the oscillation amplitudes observed experimentally.

\section{Beyond second-order perturbation theory}\label{sec:5}

So far, our analysis relies on the second-order perturbative description introduced in Eq.~(\ref{Eq:PolyKRAKEN}), which is traditionally used to interpret laser-assisted photoelectron interferometry measurements. A particular feature of highly polychromatic probe fields, such as the one employed here, is that they exhibit strong temporal intensity modulations arising from the interference between the different spectral components, as shown in Fig.~\ref{fig:6}(a). Unlike pulses with a smooth spectrum, these intensity variations occur on a timescale much shorter than the pulse envelope. For example, if the five spectral components used in this work are phase-locked and have identical peak intensities $I_0$, the instantaneous intensity reaches $I_{\mathrm{tot}}(\tau=0)=25\,I_0$ when all components interfere constructively. A few tens of femtoseconds later, however, the intensity decreases by almost two orders of magnitude before increasing again. Since the XUV pulse is only about 10\,fs long, the atom is exposed to different instantaneous IR  intensities depending on the pump-probe delay, potentially leading to deviations from the perturbative description of Eq.~(\ref{Eq:PolyKRAKEN}).

\begin{figure}[htbp]
    \centering

    % \begin{minipage}{0.48\textwidth}
    %     \centering
    %     \includegraphics[width=\textwidth]{Figures/EXP_scan_org_horizontal_50nm_III.pdf}
    % \end{minipage}
    % \hfill
    % \begin{minipage}{0.48\textwidth}
    %     \centering
    %     \includegraphics[width=\textwidth]{Figures/SFA_scan_org_horizontal_7p50e11_III.pdf}
    % \end{minipage}

    % \vspace{0.2cm}

    \begin{minipage}{\textwidth}
        \centering
        \includegraphics[width=\textwidth]{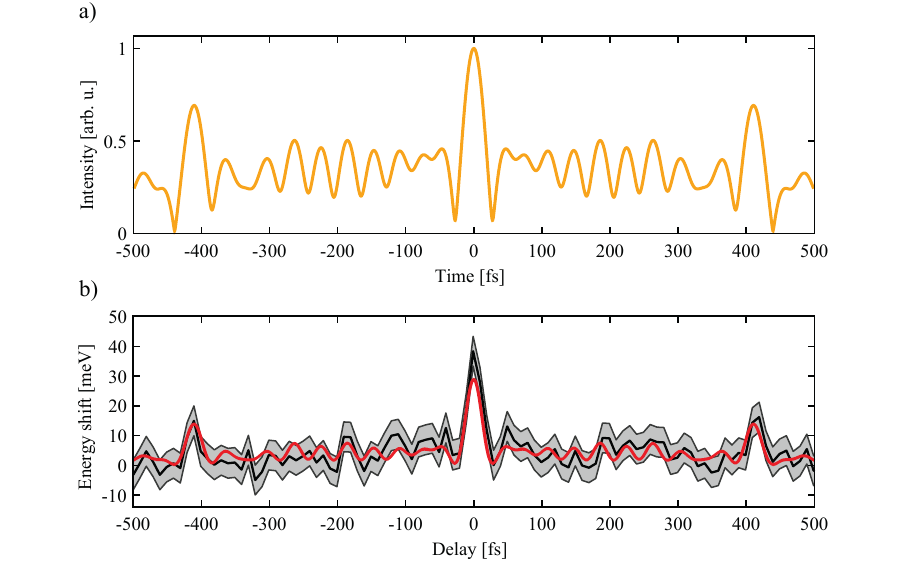}
    \end{minipage}

    \caption{a) Temporal intensity profile of the IR pulse used in SFA simulations. b) Ponderomotive shift extracted from the experimental data (black) and from the SFA calculations (red). The grey-shaded area indicates the estimated uncertainty in the experiments (10~meV).}
    \label{fig:6}
\end{figure}

The first consequence of going beyond the second-order perturbative description is the appearance of a delay-dependent ponderomotive shift of the photoelectron spectrum. To quantify this effect, we determine the center-of-mass energy of the photoelectron spectrum corresponding to absorption of harmonic 19 (not shown) as a function of pump-probe delay and reference it to the corresponding energy measured at large delays, when the XUV and IR pulses no longer overlap. Figure~\ref{fig:6}(b) shows the resulting delay-dependent energy shift (shown in black). This shift closely follows the temporal intensity profile of the IR field [Fig.~\ref{fig:6}(a)] and reaches values as large as 40 meV, corresponding to 23\% of the photoelectron peak width. The measured shift is in excellent agreement with the ponderomotive shift obtained by SFA calculations (red curve).

Beyond the delay-dependence of the ponderomotive shift, high-order multiphoton transitions can introduce distortions to the laser-assisted photoionization signal that are not accounted for in Eq.~(\ref{Eq:PolyKRAKEN}), therefore affecting the oscillation amplitudes. Within the SFA, the ponderomotive shift originates from the term proportional to $\textbf{A}^2$ in the complex exponential of Eq.~(\ref{Eq:SFA}), while the $\textbf{p}\cdot\textbf{A}$ term is responsible for continuum-continuum transitions. At sufficiently low IR intensity, the latter term gives rise to transitions via the absorption/emission of a single IR photon. In contrast, at higher intensities, transitions involving interactions with multiple IR photons become non-negligible. In the following, we isolate the contribution of the delay-dependent ponderomotive shift by comparing simulations performed with and without the $\textbf{A}^2$ term, while keeping the term linear in $\textbf{A}$. We investigate the contribution of multiphoton transitions by studying the intensity dependence of the simulations without the $\textbf{A}^2$ term.
Figure~\ref{fig:7} shows the Fourier maps obtained from SFA simulations with and without the ponderomotive term for three different IR intensities for sideband 20 (kinetic energy of $\simeq 6.5$\,eV). At the lowest intensity [$I_0=10^{10}$\,W/cm$^2$ for each spectral component, i.e. a peak intensity of $2.5\times10^{11}$\,W/cm$^2$, see Fig.~\ref{fig:7}(a,b)], the Fourier maps for both simulations are very similar, exhibiting peaks at the frequencies predicted by Eq.~(\ref{Eq:PolyKRAKEN}). The maximum value of the ponderomotive shift is only about 13\,meV, much smaller than the photoelectron spectral width, and therefore has little impact on the retrieved oscillation amplitudes. At this intensity, the results are consistent with second-order perturbation theory. 

Increasing the intensity to $I_0=5\times10^{10}$\,W/cm$^2$ per spectral component leads to a maximum ponderomotive shift of approximately 63\,meV, which represents a significant energy shift compared to the width of the photoelectron peaks. In this regime, simulations that include the $\textbf{A}^2$ term exhibit pronounced distortions in the Fourier map that are absent or reduced when the ponderomotive term is removed [compare Fig.~\ref{fig:7}(c,d)]. In particular, the  $2\,\delta\omega$ component becomes narrower in kinetic energy, while the $\delta\omega$, $3\,\delta\omega$ and $4\delta\omega$ components broaden and decrease in amplitude. In addition, weak peaks appear at frequencies corresponding to $10\,\delta\omega$ and $12\,\delta\omega$. These distortions cannot be explained by Eq.~(\ref{Eq:PolyKRAKEN}). At the highest intensity, corresponding to $I_0=10^{11}$\,W/cm$^2$ per spectral component [Fig.~\ref{fig:7}(e,f)], the ponderomotive shift can reach values as high as 125\,meV, resulting in even more pronounced distortions.

In contrast, removing the $\textbf{A}^2$ term produces Fourier maps with similar amplitudes and spectral shapes as a function of kinetic energy over the entire intensity range [Fig.~\ref{fig:7}(b,d,f)]. This demonstrates that at this kinetic energy, the intensity-dependent distortions arise primarily from the delay-dependent ponderomotive shift. The remaining difference between the lowest-intensity Fourier map [Fig.~\ref{fig:7}(b)] and the two higher intensity ones [Fig.~\ref{fig:7}(d,f)] can then be attributed to the term proportional to $\textbf{p}\cdot\textbf{A}$ in the SFA, which at large enough intensities, induces multiphoton transitions that are mostly responsible for the additional beating frequencies observed around $10\,\delta\omega$ and $12\,\delta\omega$.

\begin{figure}[htbp]
    \centering
 \includegraphics[width=\textwidth]{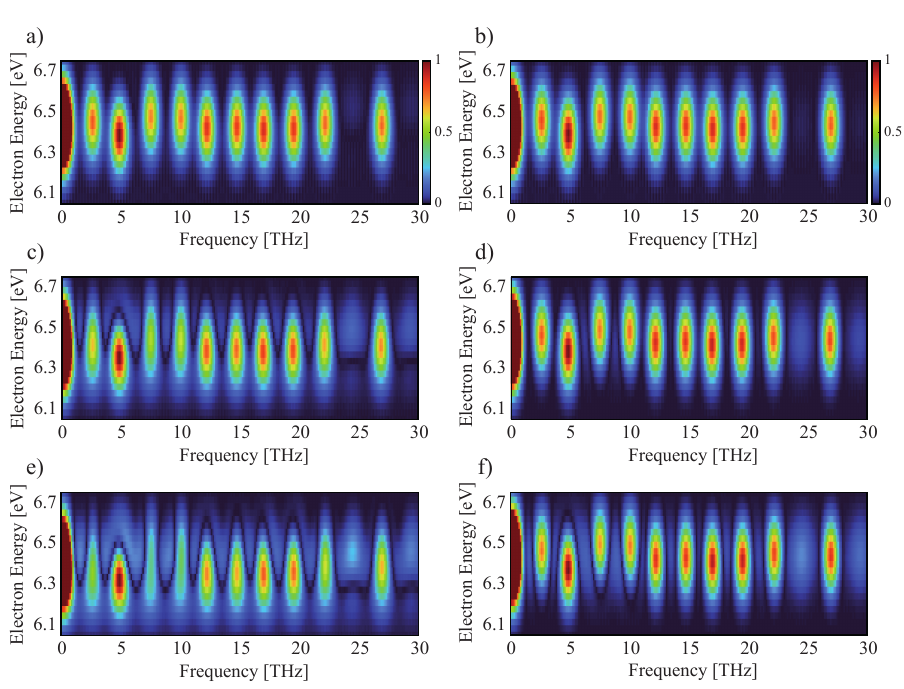}

    \caption{Normalized Fourier maps of photoelectron spectrograms obtained from SFA simulations on sideband 20 for different intensities of the individual IR spectral components: (a,b) $1.00 \times 10^{10}$\,W/cm$^2$, (c,d) $5.00 \times 10^{10}$\,W/cm$^2$, and (e,f) $1.00 \times 10^{11}$\,W/cm$^2$. (a,c,e) show simulations including the squared IR vector potential, $\textbf{A}^2$, while panels (b,d,f) show simulations without it. All Fourier maps are normalized to the peak of the $2\delta\omega$ beating frequency. A color bar is indicated on the right of (a) and (b).}
    \label{fig:7}
\end{figure}

The relative importance of the ponderomotive shift described by the $\textbf{A}^2$ term, and of the term proportional to $\textbf{p}\cdot\textbf{A}$, also depends on the photoelectron kinetic energy. Simulations for sideband 32, at an energy of approximately 25~eV demonstrate that, in this case, the Fourier maps obtained with and without the $\textbf{A}^2$ term are very similar (not shown). In addition, due to the higher kinetic energy, high-order effects originating from the $\textbf{p}\cdot\textbf{A}$ term appear at lower intensities compared to the low kinetic energy case. These calculations illustrate that the relative importance of ponderomotive shifts and higher-order laser-assisted transitions varies significantly with the photoelectron kinetic energy \cite{Maquet2007}. 
\section{Conclusions}
\label{sec:6}
In conclusion, we have experimentally demonstrated laser-assisted photoelectron interferometry using a polychromatic IR probe field composed of five spectral components forming a Golomb ruler. The resulting interferograms exhibit ten distinct beating frequencies, offering simultaneous access to multiple coherences of the photoelectron density matrix within a single scan over the relative time delay between the IR and XUV field components. We show that within second-order perturbation theory, the wavelength dependence of continuum-continuum transition amplitudes must be accounted for in polychromatic laser-assisted photoelectron interferometry. Additionally, we demonstrate that even at moderate IR intensities of the individual IR spectral components, the strongly modulated temporal profile of the polychromatic probe can introduce intensity- and delay-dependent effects that are not captured by the second-order perturbative description. First, the ponderomotive shift redistributes the spectral weights in the interferogram through a delay-dependent energy shift of the photoelectron spectrum. Second, higher-order laser-assisted transitions that involve the absorption or emission of multiple IR photons induce additional distortions. Unlike the ponderomotive shift, the importance of these higher-order processes depends strongly on the photoelectron kinetic energy.

These results establish practical guidelines for implementing polychromatic photoelectron interferometry. Operating in a perturbative regime with sufficiently weak light-matter interactions minimizes higher-order effects, although this may reduce the signal-to-noise ratio in parts of the delay scan due to the strong temporal modulations of the IR probe intensity. Retrieval algorithms that do not explicitly rely on a perturbative description of the light-matter interaction, such as mixed-FROG \cite{Bourassin2015,Bourassin2020}, may be useful for analyzing spectrograms acquired outside the perturbative regime. Alternatively, tailoring both the spectral amplitude and phase of the probe field allows reducing the peak intensity while preserving the desired spectral structure. More generally, our results suggest that careful spectro-temporal engineering of the probe field is a promising route for extending the capabilities of laser-assisted photoelectron interferometry.

\end{justify}

\ack{We thank S. Eklund and M. Barr\'e for their support during the early stages of this work. We are also grateful to C. L. M. Petersson, E. Lindroth and J. Dubois for insightful discussions.}

\funding{\begin{justify}This work was supported by the Swedish Research Council (nos. 2020-03315, 2021-04691, 2023-04603, 2023-03464, 2025-03729), LASERLAB-EUROPE (grant agreement no. 871124, European Union’s Horizon 2020 research and innovation programme), the European Research Council (Advanced Grant QPAP, 884900), the European Union's Horizon Europe research and innovation programme under the Marie Skłodowska-Curie grant agreement No 101168628 (project QU-ATTO), the CEA-Audace! Program under France 2030  (ANR24-RRII-0004-IntriQ-ATTO) and the Knut and Alice Wallenberg Foundation (grant numbers 2017.0104 and 2024.0120).  M.A., G.A., A.L. and D.B. acknowledge support from the Knut and Alice Wallenberg Foundation through the Wallenberg Centre for Quantum Technology. H.L. acknowledges support from the Swedish Research Council (2023-06502). S. L. acknowledges support from National Natural Science Foundation of China (nos. 12450402 and 12134005). C.D. acknowledges support by the Georg H. Endress foundation. C.L. acknowledges support from the French National Research Agency through grant No. ANR-24-CE29-0141-EPAD.   \end{justify}}

\roles{Conceptualization: M.A., H.L.,D.B. Data curation: E.A.B., G.A., M.A., R.W., V.S. Formal analysis: E.A.B., G.A., P.K.M., C.L. Funding acquisition: C.L., R.T., J.C., R.F., M.G., C.L.A., S.L., A.L., D.B. Investigation: E.A.B., G.A., M.A., P.K.M., C.L., R.W., V.S., H.L., M.L., H.W., C.D., M.C., C.G., R.J.S., D.B. Supervision: D.B. Writing - original draft: E.A.B., D.B. Writing - review and editing: E.A.B, G.A., M.A., P.K.M., C.L., R.W., V.S., H.L., M.L, H.W., C.D., M.C., C.G., R.T., J.C., R.J.S., R.F., M.G., C.L.A., S.L., A.L., D.B. }
 % List author names and the contributions made to the article, using terms from the NISO Contributor Roles Taxonomy (CRediT) https://credit.niso.org

%\data{Sample text inserted for demonstration.}
% For more information on IOP Publishing's research data policy see: https://publishingsupport.iopscience.iop.org/questions/research-data/

%\suppdata{Sample text inserted for demonstration.}

\bibliographystyle{iopart-num}
\bibliography{Refs}%bibliography

\end{document}